\documentclass[10pt,prx,twocolumn,amsmath,amssymb,floatfix,notitlepage,superscriptaddress]{revtex4-1}
\usepackage[T1]{fontenc}
\usepackage[utf8]{inputenc}
\usepackage{lmodern}
\usepackage{microtype,bm}
\usepackage{graphicx,booktabs}
\usepackage{times}
\usepackage[scaled=.9]{FiraSans}
\usepackage[usenames,dvipsnames]{xcolor}

\usepackage{hyperref}
\hypersetup{
  colorlinks,
  allcolors=NavyBlue,
  linktoc=all
}

\usepackage[capitalize]{cleveref}
\crefformat{section}{Sec.~#2#1#3}

\newcommand{\A}{\mathcal A}
\newcommand{\J}{\mathcal J}
\newcommand{\E}{\mathcal E}
\renewcommand*\O{\mathcal{O}} 
\newcommand*\dagg{^{\dagger}}
\newcommand*\mat[1]{\begin{pmatrix}#1\end{pmatrix}} 
\newcommand*\matr[1]{\mathsf{#1}}
\renewcommand*\vec[1]{\mathbf{#1}}

\newcommand*\up{\uparrow}
\newcommand*\down{\downarrow}
\newcommand*\ket[1]{|#1\rangle}

\DeclareMathOperator{\diag}{diag}
\DeclareMathOperator{\sgn}{sgn}

\begin{document}
\title{Topological magnon amplification}
\author{Daniel Malz}
\email{daniel.malz@mpq.mpg.de}
\affiliation{Max-Planck-Institut f\"ur Quantenoptik, Hans-Kopfermann-Strasse 1, D-85748 Garching, Germany}
\author{Johannes Knolle}
\affiliation{Blackett Laboratory, Imperial College London, London SW7 2AZ, United Kingdom}
\author{Andreas Nunnenkamp}
\affiliation{Cavendish Laboratory, University of Cambridge, Cambridge CB3 0HE, United Kingdom}
\date{\today}

\begin{abstract}
Topology is quickly becoming a cornerstone in our understanding of electronic systems.
Like their electronic counterparts, bosonic systems can exhibit a topological band structure, but in real materials it is difficult to ascertain their topological nature, as their ground state is a simple condensate or the vacuum, and one has to rely instead on excited states, for example a characteristic thermal Hall response.
Here we propose driving a topological magnon insulator with an electromagnetic field and show that this causes edge mode instabilities and a large non-equilibrium steady-state magnon edge current.
Building on this, we discuss several experimental signatures that unambiguously establish the presence of topological magnon edge modes.
Furthermore, our amplification mechanism can be employed to power a topological travelling-wave magnon amplifier and topological magnon laser, with applications in magnon spintronics.
This work thus represents a step toward functional topological magnetic materials.
\end{abstract}
\maketitle

\section{Introduction}
While fermionic topological insulators have a number of clear experimental signatures accessible through linear transport measurements~\cite{Hasan2010,Qi2011}, noninteracting bosonic systems with topological band structure have a simple condensate or the vacuum as their ground state~\cite{Vishwanath2013}, making it more difficult to ascertain their topological nature.
Their excited states, however, may carry signatures of the to\-po\-lo\-gy of the band structure, for example in form of a thermal Hall response~\cite{Katsura2010,Onose2010,Hirschberger2015}.
There is great interest in certifying and exploiting topological edge modes in bosonic systems, as they are chiral and robust against disorder, 
making them a great resource to realize backscattering-free waveguides~\cite{Haldane2008,Wang2009}
and potentially topologically protected travelling-wave amplifiers~\cite{Peano2016}.
It has been predicted that topological magnon insulators (TMI) are realized, e.g.,
in kagome planes of certain pyrochlore magnetic insulators as a result of Dzyaloshinskii-Moriya (DM) interaction~\cite{Katsura2010,Zhang2013,Mook2014}.
To date, there exists only indirect experimental proof, via neutron scattering measurements of the bulk band structure in Cu[1,3-benzenedicarboxylate (bdc)]~\cite{Chisnell2015},
and observation of a thermal magnon Hall effect in Lu$_2$V$_2$O$_7$~\cite{Onose2010} and Cu(1,3-bdc)~\cite{Hirschberger2015}.
The main obstacle is that magnons are uncharged excitations and thus invisible to experimental tools like STM or ARPES with spatial resolution.
An unambiguous experimental signature, such as the direct observation of an edge mode in the bulk gap is hampered by limitations
in energy resolution (in resonant x-ray scattering) or signal strength (in neutron scattering)~\cite{Chisnell2015}.

Here, we propose driving a magnon edge mode to a parametric instability, which, when taking into account nonlinear damping, induces a non-equilibrium steady state with a large chiral edge mode population.
In such a state the local polarization and magnetization associated with the edge mode are coherently enhanced, which could enable direct detection of edge modes via neutron scattering.
Crucially, we show that selective amplification of edge modes can be achieved while preserving the stability of the bulk modes and thus the magnetic order.
Another key experimental signature we predict is that applying a driving field gradient gives rise to a temperature gradient along the transverse direction, thus establishing what one might call a driven Hall effect (DHE).
Topological magnon amplification has further uses in magnon spintronics~\cite{Chumak2015}, providing a way to amplify magnons and to build a topological magnon laser~\cite{Harari2018,Bandres2018}.
Our work on driving topological edge modes in magnetic materials complements previous investigations in ultracold gases~\cite{Galilo2015,Galilo2017}, photonic crystals~\cite{Peano2016}, and most recently arrays of semiconductor microresonators~\cite{Harari2018,Bandres2018} and graphene~\cite{Plank2018}.

\section{Results}
\subsection{Edge mode parametric instability}\label{sec:instability}
\begin{figure}[tb]
  \centering
  \includegraphics[width=\linewidth]{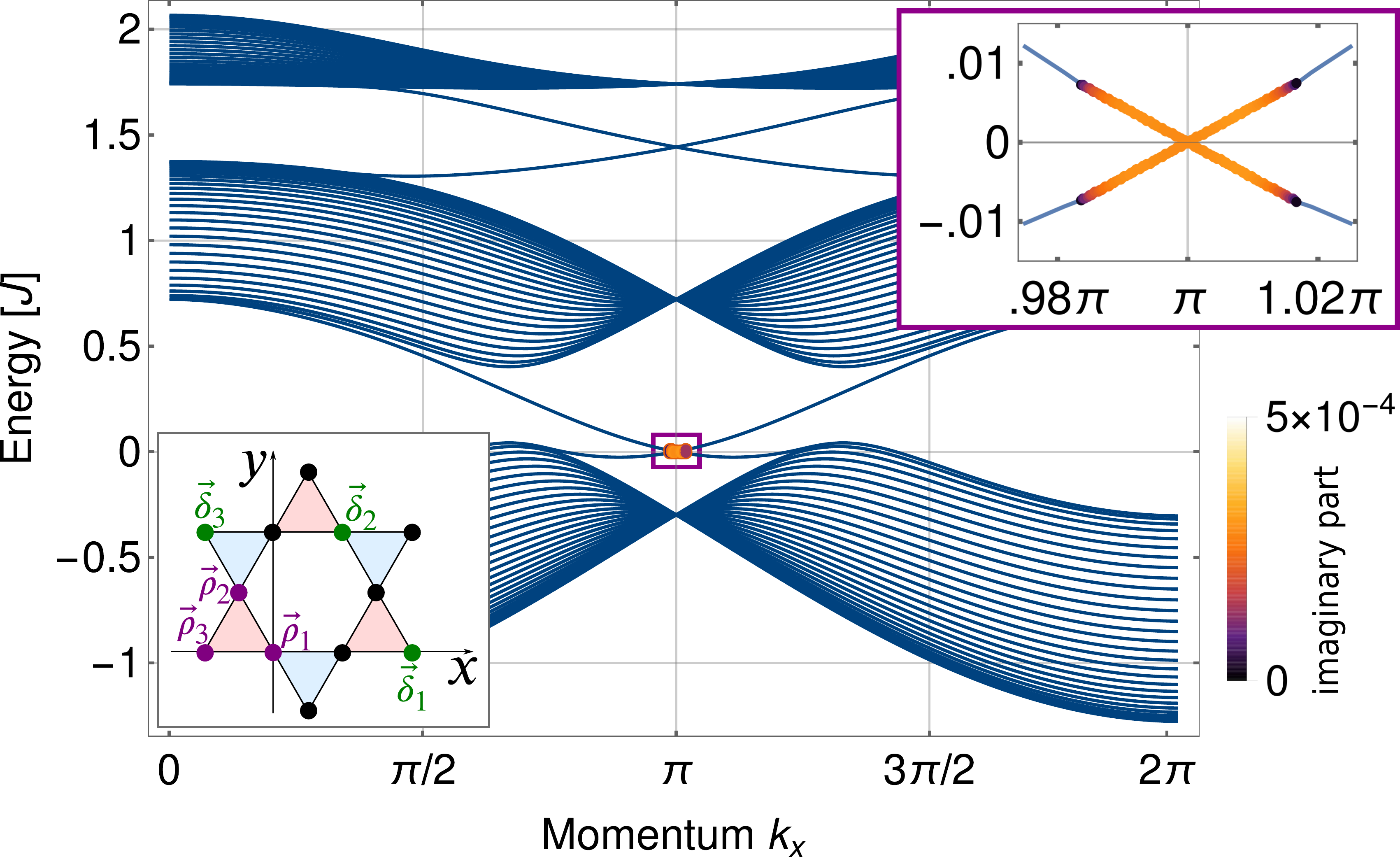}
  \caption{\textbf{Magnon band structure and instability.} The band structure of a kagome topological magnetic insulator (TMI) strip with a drive tuned to the edge modes at momentum $\pi$ (lattice constant $a=1$), calculated from the microscopic model in \cref{sec:model}.
  Instabilities are marked by coloured disks, with the colour representing the size of the imaginary part, in units of $J$.
  Bottom left (grey) inset: A sketch of a few unit cells of the kagome lattice, with lattice vectors $\bm\delta_i$ ($|\bm\delta_i|=a=1$) and site vectors $\bm\rho_i$ drawn in.
  The system we consider is infinite along the $x$-direction and comprises $W=45$ unit cells along the $y$-direction.
  Top right (purple) inset: Zoom of the unstable region.
  Parameters: $D_z/J=0.2$, $h/J=0.01$, $\Omega_0/J=2.578$, $\hat{\vec e}=(0,1)$,
  $\E/J=0.0004$, $\gamma/J=0.0001$.
}
\label{fig:instabilities}
\end{figure}
Before discussing a microscopic model, we show qualitatively how a parametric instability may arise from anomalous magnon pairing terms in a chiral one-dimensional wave\-guide.
We consider bosonic modes $\{\hat a_k\}$ with energies $\{\omega_k\}$ (for example the magnon edge mode between the first and second band, as in \cref{fig:instabilities}), labelled by momentum $k$, interacting with another bosonic mode (electromagnetic field mode) $\hat b$, as described by generic three-wave mixing ($\hbar=1$)
\begin{equation}
  \hat H_{\mathrm{int}}=\sum_k\frac{g_k}{2}\left(\hat a_{-k}\dagg \hat a_{k}\dagg \hat b+\hat b\dagg \hat a_{k}\hat a_{-k}\right).
  \label{eq:three_wave_mixing}
\end{equation}
Under strong coherent driving, the bosonic annihilation operator $\hat b$ can be replaced by its classical amplitude $\hat b\approx\beta\exp(-i\Omega_0t)\gg1$,
yielding an effective Hamiltonian
$\hat H=\hat H_0+g_k\beta[\hat a_{-k}\dagg \hat a_k\dagg \exp(i\Omega_0t)+\hat a_k\hat a_{-k}\exp(-i\Omega_0t)]$.
The second term produces magnon pairs with equal and opposite momentum.
The time-dependence can be removed by passing to a rotating frame with respect to $\sum_k(\Omega_0/2)\hat a_k\dagg \hat a_k$.
From the Hamiltonian it is straightforward to derive the equations of motion, which couple particles at momentum $k$ with holes at $-k$.
Neglecting fluctuations, we focus on the classical amplitudes of the fields $\alpha_k=\langle \hat a_k\rangle$ and include a phenomenological linear damping rate $\gamma_k$ to take into account the various damping processes present in such materials~\cite{Chisnell2015,Chernyshev2016}.
As we are interested in amplification around a small bandwidth, we neglect the momentum dependence of the coupling $g_k\simeq g$ and damping $\gamma_k\simeq \gamma$, arriving at
\begin{equation}
  i\frac{d}{dt}\vec{A}_k
  =\mat{\tilde\omega_k-i\frac{\gamma}{2}& \E \\ -\E & -\tilde\omega_{-k}-i\frac{\gamma}{2}}\vec{A}_k,
  \label{eq:two_mode_eom}
\end{equation}
where we have introduced the frequency relative to the rotating frame $\tilde\omega_k=\omega_k-\Omega_0/2$, the vector $\vec A_k=(\alpha_k, \alpha_{-k}^*)$, and the overall coupling strength $\E=g\beta$.
The eigenvalues of the dynamical matrix \cref{eq:two_mode_eom} are the complex energies
\begin{equation}
  \omega_{k,\pm}=\frac{\omega_k-\omega_{-k}}{2}-\frac{i\gamma}{2}\pm \sqrt{\frac{(\tilde\omega_k+\tilde\omega_{-k})^2}{4}-\E}.
  \label{eq:two_mode_evals}
\end{equation}
If the coupling $\E$ exceeds the energy difference between pump photons and magnon pair (the detuning) $\tilde\omega_k+\tilde\omega_{-k}=\omega_k+\omega_{-k}-\Omega_0$, the square root becomes imaginary. 
If further its magnitude exceeds $\gamma$, more particles are created than dissipated, causing an instability and exponential growth of the number of particles in this mode. Eventually the growth is limited by nonlinear effects, as discussed below.
Despite its simplicity, \cref{eq:two_mode_evals} provides an accurate account of the fundamental instability mechanism in two-dimensional kagome TMIs, as is illustrated by the quantitative agreement \cref{fig:edge_instability}.
This forms the key ingredient for directly observing chiral magnon edge modes.

\subsection{Microscopic model}\label{sec:model}

Turning to a more realistic model, we consider spins on the vertices of an insulating kagome lattice ferromagnet that interact via Heisenberg and Dzyaloshinskii-Moriya (DM) interaction
\begin{equation}
  \hat H_0=\sum_{\langle ij\rangle }\left[ -J \hat{\vec S}_i\cdot\hat{\vec S}_j+\vec D_{ij}\cdot(\hat{\vec S}_i\times \hat{\vec S}_j) \right]-g_{\mathrm{L}}\mu_{\mathrm{B}} \vec H_0\cdot\sum_i\hat{\vec S}_i.
  \label{eq:bare_hamiltonian}
\end{equation}
Here, $\vec D_{ij}$ is the DM vector that can in principle differ from bond to bond, but is heavily constrained by lattice symmetries. $\vec H_0$ is an externally applied magnetic field, $\mu_{\mathrm{B}}$ the Bohr magneton, $g_{\mathrm{L}}$ the Land\'e g-factor, and $J$ is the Heisenberg interaction strength. 
This model has been found to describe the thermal magnon Hall effect in Lu$_2$V$_2$O$_7$~\cite{Onose2010},
as well as the bulk magnon band structure of Cu(1,3-bdc)~\cite{Chisnell2015}.

The low energy excitations around the ferromagnetic order are magnons, whose bilinear Hamiltonian
is obtained from a standard Holstein-Primakoff transformation to order $1/S$ along the direction of magnetization, \emph{i.e.}, $\hat S^+=\sqrt{2s}\,\hat a,\hat S^-=\sqrt{2s}\,\hat a\dagg,\hat S^z=s-\hat a\dagg \hat a$~\cite{Katsura2010,Zhang2013}, yielding
\begin{equation}
  \hat H_0=-\frac12(J+iD_z)\sum_{\langle mn\rangle }\hat a_m\dagg\hat a_n+\text{H.c.}+h\sum_m\hat a_m\dagg\hat a_m+
  K_0,
  \label{eq:bilinear_magnon_hamiltonian}
\end{equation}
where $K_0$ is a constant, the sum ranges over bonds directed counterclockwise in each triangle, and we have chosen the magnetic field to point along $z$, introducing $h\equiv g_{\mathrm{L}}\mu_{\mathrm{B}}H_0^z$.

To second order, the Hamiltonian only contains the component of $\vec D_{ij}$ along $z$ ($D_z$), which is the same for all bonds due to symmetry.
We take the unit cell to be one upright triangle (red in \cref{fig:instabilities}),
with sites $\bm\rho_1=(0,0)$, $\bm\rho_2=(-1,\sqrt{3})/4$, $\bm\rho_3=(-1/2,0)$.
The unit cells form a triangular Bravais lattice generated by the lattice vectors
$\bm\delta_1=(1,0)$, $\bm\delta_2=(1,\sqrt{3})/2$, $\bm\delta_3=\bm\delta_2-\bm\delta_1=(-1,\sqrt{3})/2$.
For nonzero $D_z$, the bands in this model are topological~\cite{Zhang2013,Chisnell2015}
causing exponentially localized edge modes to appear within the band gaps.

The effect of an oscillating electric field on magnons in a TMI is characterized by the polarization operator, 
which can be expanded as a sum of single-spin terms, products of two spins, three spins, etc.~\cite{Moriya1968}
Lattice symmetries restrict which terms may appear in the polarization tensor~\cite{Moriya1968}.
In the pyrochlore lattice, the polarization due to single spins (linear Stark effect) vanishes, as each lattice site is a centre of inversion,
such that the leading term contains two spin operators.
The associated tensor can be decomposed into the isotropic (trace) part $\bm\pi$, as well as the anisotropic traceless symmetric and antisymmetric parts $\bm\Gamma$ and $\vec D$, \emph{viz.}
$\hat{\vec P}_{jl}=(\bm\pi_{jl}\delta^{\beta\gamma}
+\vec\Gamma_{jl}^{(\beta\gamma)}+\vec D_{jl}^{[\beta\gamma]})
\hat S_{j}^\beta\hat S_{l}^\gamma$ (sum over $\beta,\gamma$ implied). 
Kagome TMIs generically have a nonzero anisotropic symmetric part, which implies the presence of anomalous magnon pairing terms in the spin-wave picture 
\begin{equation}
  \begin{aligned}
  	\hat{\vec P}_{mn}&=\left(\bm\Gamma_{mn}^{\alpha,xx}-\bm\Gamma_{mn}^{\alpha,yy}-2i\bm\Gamma_{mn}^{\alpha,(xy)} \right)\hat a_m\hat a_n+\cdots\\
  	&\equiv\vec Q_{mn}\hat a_m\hat a_n+\text{H.c.}+\cdots.
  \end{aligned}
  \label{eq:anomalous_terms_in_polarization}
\end{equation}
The polarization enters the Hamiltonian via coupling to the amplitude of the electric field,
$\hat H(t)=\hat H_0-\vec E(t)\cdot\hat{\vec P}$, thus introducing terms that create a pair of magnons while absorbing a photon.
Pair production of magnons is a generic feature of antiferromagnets (via $\bm\pi$),~\cite{Moriya1968}
but since in ferromagnets it relies on anisotropy, it is expected to be considerably weaker.
A microscopic calculation based on a third-order hopping process in the Fermi-Hubbard model at half filling reveals that $|\vec Q|=ae(t/U)^3/2$,
where $a$ is the lattice vector, $e$ the elementary charge, $t$ the hopping amplitude, and $U$ the on-site repulsion (see Supplementary Note 1).

As in the chiral waveguide model, assume an oscillating electric field $\vec E(t)=\vec E_0\cos(\Omega_0t)$.
We consider an infinite strip with $W$ unit cells along $y$, but remove the lowest row of sites to obtain a manifestly inversion-symmetric model.
Diagonalizing the undriven Hamiltonian $\hat H_0$ \eqref{eq:bare_hamiltonian}, we label the eigenstates $b_{k,s}$ by their momentum along $x$ and an index $s\in\{1,2,\cdots,3W-2\}$.
After performing the rotating-wave approximation, the full Hamiltonian reads
\begin{equation}
  \hat H=\sum_{k,s}\tilde\omega_{k,s}\hat b_{k,s}\dagg \hat b_{k,s}
  -\frac{1}{2}\left[\vec E_0\cdot \vec{\tilde Q}_{ss'}(k)\hat b_{k,s}\hat b_{-k,s'}+\text{H.c.}\right],
  \label{eq:full_hamiltonian}
\end{equation}
where we have introduced
$\tilde \omega_{k,s}=\omega_{k,s}-\Omega_0/2$, and $\vec{\tilde Q}_{ss'}(k)$, which characterizes the strength of the anomalous coupling between two modes.
It is obtained from $\vec{Q}_{mn}$ through Fourier transform and rotation into the energy eigenbasis (\emph{cf.}~Supplementary Note 3).

\begin{figure}[tb]
  \centering
  \includegraphics[width=\linewidth]{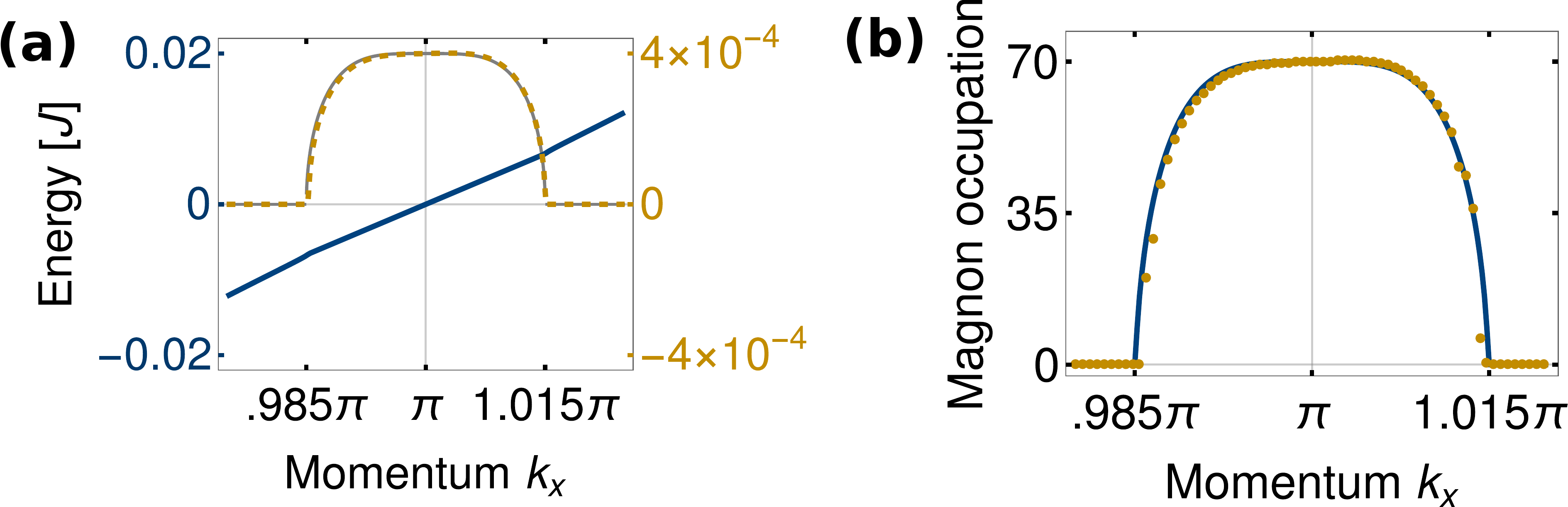}
  \caption{\textbf{Characterization of instability.} \textbf{(a)} Complex energy of the edge eigenmodes (\emph{cf.}~\cref{eq:two_mode_evals}).
  The real part (frequency) is plotted in blue, whereas the imaginary part (amplification) is shown in yellow.
  The perfectly matching grey curve is the theory \cref{eq:two_mode_evals}
  with $\E=g_kE_0=|\vec E_0| |\vec{\tilde Q}_{\bar s\bar s}(\pi)|=4\times10^{-4}J$
  and $\omega_\pi''\simeq0.3605J$ (numerically extracted from band structure, not fitted to instability).
  Note that the excellent agreement only holds if the polarization of the electric field points along $y$, \emph{i.e.}, along the width of the strip, as we explain in more detail in Supplementary Note 4.
  \textbf{(b)} The steady-state edge occupation calculated from \cref{eq:nonlinear_eom} with the same parameters as in \cref{fig:instabilities} and nonlinear damping $\eta=10^{-5}J$ (the same in microscopic theory and chiral waveguide model).
	Shown in blue is the chiral waveguide model, the yellow dots are calculated numerically from the microscopic Hamiltonian.
   Parameters are as in \cref{fig:instabilities}, but with $W=15$.
}
\label{fig:edge_instability}
\end{figure}

As in the one-dimensional waveguide model, a pair of modes is rendered unstable if their detuning $\Delta_{k,ss'}=\omega_{k,s}+\omega_{-k,s'}-\Omega_0$ is smaller than the anomalous coupling between them.
The detuning $\Delta_{k,ss'}$ varies quickly as a function of $k$ except at points where the slopes of $\omega_{k,s}$ and $\omega_{-k,s'}$ coincide to first order, which happens at $k=0,\pi$ when $s=s'$.
At those values of $k$, the energy matching condition is fulfilled for a broader range of wavevectors, which leads to a larger amplification bandwidth.
However, the edge modes are only localized to the edge around $k=\pi$, such that driving around $k=\pi$ is most efficient, which we consider here (\emph{cf.}~\cref{fig:instabilities}).
Expanding the dispersion to second order around this point, we find
$\omega_{\pi+q}\simeq \omega_\pi+q\omega_\pi'+(q^2/2)\omega_\pi''+\O(q^3)$, yielding
$\Delta_{\pi+q}=2\omega_\pi-\Omega_0+q^2\omega_\pi''+\O(q^4)$.
Placing the pump at $\Omega_0=2\omega_{\pi}$
thus makes magnon pairs around $k=\pi$ resonant, on a bandwidth of order $\sqrt{\E/\omega_\pi''}$.
For weak driving, where the bandwidth is low, higher-order terms in the dispersion relation can be neglected, and this simple calculation captures the amplification behaviour extremely well, as we illustrate in \cref{fig:edge_instability}a.
We calculate the band structure and find the unstable modes numerically (see Methods), with $\tilde\omega_{k,s}$ and $\vec{\tilde Q}_{ss'}$ obtained from a microscopic model detailed in Supplementary Note 2, and plot the resulting band structure with instabilities in \cref{fig:instabilities}.

As we have seen above, an instability requires the anomalous terms to overcome the linear damping and the effective detuning.
Linear damping, which we include as a uniform phenomenological parameter $\gamma$, has important consequences, as it sets a lower bound for the amplitude of the electrical field required to drive the system to an instability. It also ensures bulk stability.
We have seen that there are three conditions for a parametric instability. First, there has to exist a pair of modes whose lattice momenta add to 0 (or $2\pi$); Second, the sum of their energy has to match the pump frequency; and Third, the strength of their anomalous interaction has to overcome both their detuning and their damping. 
While momentum and energy matching is by design fulfilled by the edge mode, there is a large number of bulk mode pairs that also fulfil it. 
We show in Supplementary Note 4 that choosing the polarization to lie along $y$ increases the anomalous coupling for modes with wavevector close to $\pi$
and that the coupling is small for almost all bulk mode pairs.
The reason for this is that the bulk modes are approximately standing waves along $y$, and most bulk mode pairs have differing numbers of nodes, such that their overlap averages to zero.
The remaining modes with appreciable anomalous coupling a far detuned in energy. 
This way, robust edge state instability can be achieved without any bulk instabilities, as demonstrated in \cref{fig:instabilities}.
Bulk stability is crucial for the validity of the following discussion.

\begin{figure*}[tb]
  \centering
  \includegraphics[width=\linewidth]{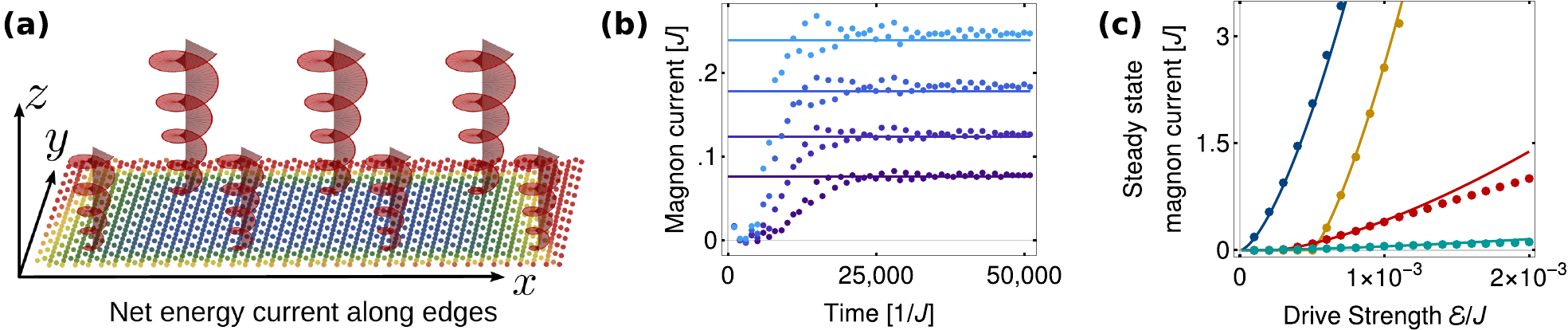}
  \caption{
  	\textbf{Driving and resulting edge current.}
  	\textbf{(a)} The two-dimensional kagome lattice ferromagnet is driven by an electromagnetic field perpendicular to the kagome plane. 
  	To observe the driven Hall effect (DHE), the field is applied with a linear gradient along $y$, which leads to a temperature difference along $x$. Colour gradient from blue to red indicates increasing temperature.
	\textbf{(b)} The current as function of time.
	The shades of blue from dark to bright correspond to $\E=\{3,4,5,6\}\times10^{-4}J$, respectively.
	The straight line is the theoretical prediction for the steady-state value~\eqref{eq:JSS}. 
	From the equation of motion one can estimate the time to reach the steady state to be of order $t_{\mathrm{eq}}\sim \E^{-1}\log(\E/\eta)$.
	\textbf{(c)} The steady-state particle current plotted against the drive strength $\E$.
	The solid lines correspond to the theoretical formula \cref{eq:JSS}, whereas the dots are calculated numerically.
	In $\{$blue, yellow, red, turquoise$\}$ we show $(\gamma,\eta)=\{(1,0.1),(100,0.1),(50,1),(10,10)\}\times10^{-5}J$.
	We see that \cref{eq:JSS} agrees well with the numerically calculated steady-state current.
	The yellow, red, and turquoise curves have kinks at $2\E=\gamma$, which mark the onset of instability. 
	The clear deviation occurs once bulk modes become unstable, in which case our approximations break down and system ceases to remain close to the ground state.
	All unspecified parameters are as in \cref{fig:instabilities}, except for $W=15$.
  }
  \label{fig:current}
\end{figure*}

In the presence of an instability, the linear theory predicts exponential growth of edge magnon population.
In a real system, the exponential growth is limited by nonlinear damping, for which we introduce another uniform parameter $\eta$,
in the same spirit as Gilbert damping in nonlinear Landau-Lifshitz-Gilbert equations~\cite{Ruckriegel2018}, such that \cref{eq:two_mode_eom} becomes
\begin{equation}
  i\frac{d}{dt}\vec{A}_k
  =\mat{\tilde\omega_k-i\frac{\gamma+\eta|\alpha_k|^2}{2}& \E \\ -\E & -\tilde\omega_{-k}-i\frac{\gamma+\eta|\alpha_k|^2}{2}}\vec{A}_k.
  \label{eq:nonlinear_eom}
\end{equation}
Microscopically, such damping arises from the next order in the spin-wave expansion that allows four-wave mixing.
While the linear theory only predicts the instability, \cref{eq:nonlinear_eom} predicts a steady-state magnon occupation given through
$|\alpha_{\pi+q}|^2=\eta^{-1}(\sqrt{4\E^2-q^4(\omega_\pi'')^2}-\gamma)$ (\emph{cf.}~Methods), which we show in \cref{fig:edge_instability}b.

\subsection{Experimental signatures}

TMIs exhibit a magnonic thermal Hall effect at low temperatures~\cite{Katsura2010,Onose2010}.
A similar effect occurs when the magnon population is not thermal, but a consequence of coherent driving, realizing a driven Hall effect (DHE).

We can calculate the steady-state edge magnon current from the occupation calculated above,
\begin{equation}
  J_{\mathrm{SS}}=\int_{-\Lambda}^\Lambda\frac{dq}{2\pi}\,|\alpha_{\pi+q}|^2\nu_{\pi+q},
  \label{eq:JSS_formula}
\end{equation}
where $\nu_{\pi+q}\simeq\omega_{\pi}'+q\omega_{\pi}''$ is the group velocity and $\Lambda=\sqrt[4]{(4\E^2-\gamma^2)/(\omega_{\pi}'')^2}$ is the range over which the steady-state population is finite (which coincides with the range over which the modes become unstable).
While the integral can be done exactly (\emph{cf.}~Methods)
an approximation within $\pm5\%$ is given through
\begin{equation}
  J_{\mathrm{SS}}(\E)\approx \frac{5\omega_\pi'}{6\eta\pi}\left(\frac{4\E^2-\gamma^2}{(\omega_{\pi}'')^2}\right)^{\tfrac14}(2\E-\gamma).
  \label{eq:JSS}
\end{equation}
For $2\E\gg\gamma$, a characteristic scaling of steady-state current with driving strength appears, $J_{\mathrm{SS}}\propto \E^{3/2}$, distinct from the linear dependence one would expect for standard heating.

In \cref{fig:current}b,c, we demonstrate that the steady-state edge current depends on the drive strength in a fashion that is well described by \cref{eq:JSS}.
The order-of-magnitude equilibration time can be estimated from the solution to $\dot\alpha=(1/2)(\E-\eta|\alpha|^2)\alpha$, and for $\eta/\E\gg1$ it evaluates to $t_{\mathrm{eq}}\sim \E^{-1}\log(\E/\eta)\sim 10^4J^{-1}$ for our chosen values of $\E$ and $\eta$. 

A DHE arises when a rectangular slab of size $L_x\times L_y$ is driven by a field with a gradient along $y$, as sketched in \cref{fig:current}a.
If $L_x,L_y\gg \nu_\pi t_{\mathrm{eq}}$, the edges equilibrate to a steady-state magnon population governed by \cref{eq:JSS}.
The difference between the steady-state magnon currents on top and bottom edge corresponds to a
net energy current $J_{\mathrm{net}}^x$ along $x$, which to first order in the drive strength difference $\Delta_y\E$ can be written~\cite{Vinkler-Aviv2018}
\begin{equation}
  J_{\mathrm{net}}^x=\kappa_{xy}(\E_{\mathrm{avg}})\Delta_y \E,\qquad
  \kappa_{xy}(\E)\equiv\frac{dJ_{\mathrm{SS}}(\E)}{d\E}
  \label{eq:net_energy_current}
\end{equation}
where one should note that in this non-equilibrium setting $\kappa_{xy}$ is not a proper conductivity as in conventional linear response.
The net edge current causes one side of the system to heat up faster, resulting in a temperature difference transverse to the gradient.
As the edge magnons decay along the edge, the reverse heat current is carried by bulk modes.
For small temperature differences the heat current follows the temperature gradient linearly and thus
the averaged temperature difference $\Delta_x T=\int dy\,[T(L_x,y)-T(0,y)]/L_y=J_{\mathrm{net}}^x/\kappa_{xx}$.
The temperature difference can thus be written in terms of the applied field strength difference
\begin{equation}
  \Delta_x T=\frac{\Delta_y\E}{\kappa_{xx}}\kappa_{xy}(\E).
  \label{eq:Hall_coefficient}
\end{equation}
As a word of caution, we note that this relation relies on several key assumptions. 
To begin with, temperature is in fact not well defined along the edge, as there is a non-equilibrium magnon occupation. 
Edge magnons decay at a certain rate into phonons, which can be modelled as heating of the phonon bath.
If the equilibration time scale of the latter is fast compared to the heating rate through magnon decay, one can at least associate a local temperature to the phonons. 
Similarly, the bulk magnon modes can be viewed as a fast bath for the magnon edge mode and similar considerations apply.
Even if these assumptions are justified, the two baths do not need to have the same temperature. 
Next, the heat conductivity associated to magnons and phonons differ in general, such that the $\kappa_{xx}$ appearing in \cref{eq:Hall_coefficient} can only be associated with the bulk heat conductivity if the temperatures of the two baths are equal. 
Some of these complications have been recognized to also play an important role in measurements of the magnon thermal Hall effect~\cite{Vinkler-Aviv2018}.

While the above-mentioned concerns make quantitative predictions difficult, the DHE is easily distinguishable from the thermal Hall effect, due to the strong dependence of the temperature difference $\Delta_xT$ on drive frequency and polarization,
as well as the fact that below the cutoff $2\E=\gamma$ no instability occurs and that $J_{\mathrm{net}}\propto \sqrt{\E_{\mathrm{avg}}}$ for $2\E_{\mathrm{avg}}\gg\gamma$, rather than the linear dependence one would expect from standard heating.
In certain materials such as Cu(1,3-bdc), the appearance or disappearance of the topological edge modes can be tuned with an applied magnetic field~\cite{Chisnell2015},
a property that could be used to further corroborate the results of such an experiment.

A number of other experimental probes might be used to certify a large edge magnon current and thus the presence of edge states. 
On the one hand, with a large coherent magnon population in a given mode, the local magnetic field and electric polarization associated to that mode will be enhanced.
In particular techniques that directly probe local magnetic or electric fields, such as neutron scattering~\cite{Chisnell2015,Yao2018} or x-ray scattering,
which to date are not powerful enough to resolve edge modes~\cite{Chisnell2015}, would thus have a coherently enhanced signal, for example by almost two orders of magnitude when taking the conservative parameters in \cref{fig:edge_instability}.
On the other hand, heterostructures provide a way to couple the magnons out of the edge mode into another material~\cite{Ruckriegel2018}, for example one with a strong spin Hall effect, in which they can be detected more easily.
In this setup, again the fact that the edge magnons have a large coherent population should make their signal easily distinguishable from thermal noise.

\subsection{Material realizations}

The model of a kagome lattice ferromagnet with DM interaction has been found to describe the thermal magnon Hall effect in Lu$_2$V$_2$O$_7$~\cite{Onose2010}, as well as the bulk magnon band structure of Cu(1,3-bdc)~\cite{Nytko2008,Chisnell2015}.
These materials are in fact 3D pyrochlore lattices, which can be pictured as alternating kagome and triangular lattices along the [111] direction.
However, their topological properties can be captured by considering only the kagome planes~\cite{Zhang2013,Mook2014,Chisnell2015} (shown in \cref{fig:instabilities}), thus neglecting the coupling between kagome and triangular planes.
It has been suggested that the effect of the interaction may be subsumed into new effective interaction strengths~\cite{Mook2014} or into an effective on-site potential~\cite{Zhang2013}.
Typical values for strength of the DM and Heisenberg interactions lie between $|\vec D|/J\approx 0.18$~\cite{Chisnell2015}, $J\approx0.6\pm0.1\,\text{meV}\simeq 150\pm30\,\text{GHz}$~\cite{Chisnell2015} in Cu(1,3-bdc) and $|\vec D|/J\approx0.32$~\cite{Onose2010}, $J\approx 3.4\,$meV$\,\approx 0.82\,$THz~\cite{Zhang2013} in Lu$_2$V$_2$O$_7$.
The energy of the edge states close to $k=\pi$ is approximately $J$, such that the applied drive needs to be at a frequency $\omega_0/2\pi=0.3$--$1.6\,$THz.
While experimentally challenging, low THz driving down to 0.6 THz has recently been achieved~\cite{Karch2011,Plank2018}.
Furthermore, the magnon energy can be tuned by applied magnetic fields. 

An instability requires $E_0ae(t/U)^3\gtrsim \gamma$.
With $a\simeq 10\,$\AA~\cite{Zhang2013}, $J\simeq1\,$meV, and assuming $t/U\simeq0.1$, we can estimate the minimum field strength required to overcome damping $\gamma_k\simeq10^{-4}J$ to be $E_0\simeq10^5\,$V/m, although for quantitative estimates one would require both accurate values for the damping of the edge modes (at zero temperature) and $t/U$. 
This is accessible in pulsed operation~\cite{Karch2011,Plank2018}, and perhaps in continuous operation through the assistance of a cavity.

Since the qualitative behaviour we describe can be derived from general and phenomenological considerations, we expect it to be robust and present in a range of systems, as long as they allow for anisotropy, i.e., if bonds are not centres of inversion.
We thus expect that topological magnon amplification is also possible in recently discovered topological honeycomb ferromagnet CrI$_3$~\cite{Chen2018}.

\section{Discussion}\label{sec:conclusions}
We have shown that appropriate electromagnetic driving can render topological magnon edge modes unstable, while leaving the bulk modes stable.
The resulting non-equilibrium steady state has a macroscopic edge magnon population.
We present several strategies to certify the topological nature of the band structure, namely, implementing a driven Hall effect (DHE), direct detection with neutron scattering, or by coupling the magnons into a material with a spin Hall effect.

Our work paves the way for a number of future studies.
As we have pointed out, edge mode damping plays an important role here.
One might expect their damping to be smaller than that of generic bulk modes as due to their localization they have a smaller overlap to bulk modes.
This suppression should be compounded by the effect of disorder~\cite{Ruckriegel2018}, which may further enhance the feasibility of our proposed experiments.
On the other hand, rough edges will have an influence over the matrix element between drive and edge modes, leading to variations in the anomalous coupling strength.
Phonons in the material are crucial for robust thermal Hall measurements~\cite{Vinkler-Aviv2018} and could possibly mix with the chiral magnon mode~\cite{Thingstad2018},
which motivates full microscopic calculations.

An exciting prospect is to use topological magnon amplification in magnon spintronics.
There have already been theoretical efforts studying how magnons can be injected into topological edge modes with the inverse spin Hall effect~\cite{Ruckriegel2018}.
Given an efficient mechanism to couple magnons into and out of the edge modes,
our amplification mechanism may enable chiral travelling-wave magnon amplifiers,
initially proposed in photonic crystals~\cite{Peano2016}.
Even when simply seeded by thermal or quantum fluctuations, the large coherent magnon steady state could power topological magnon lasers~\cite{Harari2018},
with tremendous promise for future application in spintronics.
In the near future, we hope that topological magnon amplification can be used for an unambiguous discovery of topological magnon edge modes.

\begin{acknowledgments}
	We are grateful to Ryan Barnett, Derek Lee, Rub\'en Otxoa, Pierre Roy, and Koji Usami for insightful discussions and helpful comments.
	DM acknowledges support by the Horizon 2020 ERC Advanced Grant QUENOCOBA (grant agreement 742102). 
	AN holds a University Research Fellowship from the Royal Society and acknowledges support from the Winton Programme for the Physics of Sustainability
	and the European Union's Horizon 2020 research and innovation programme under grant agreement No 732894 (FET Proactive HOT).
\end{acknowledgments}

\section*{Methods}
  \subsection*{Numerical Calculation}For the numerical calculation, we choose a manifestly inversion-symmetric system obtained by deleting the lowest row of sites,
  a situation that is depicted in \cref{fig:instabilities}, where the tip of the lowest blue triangle is part of a unit cell whose other sites are not included.
  For example, repeating the star shown in \cref{fig:instabilities} along $x$ would result in an inversion-symmetric strip with $W=3$.
  A Fourier transform of \cref{eq:bilinear_magnon_hamiltonian} along $x$ yields a $3W-2$ by $3W-2$ Hamiltonian matrix for each momentum $k$ 
  \begin{equation}
  	\begin{aligned}
	  &H_0=K_0
	  -\frac{1}{2}\left\{ (J+iD_z)\sum_{k, l_y}
	  \left[
	   	 e^{-ik/4}a_{1,k,l_y}\dagg a_{2,k,l_y}\right.\right.\\
	   &+e^{-ik/4}a_{2,k,l_y}\dagg a_{3,k,l_y}
	  	+2\cos(k/2)a_{3,k,l_y}\dagg a_{1,k,l_y} \\
	  	&\left.\left.
  	  	+e^{ik/4}a_{1,k,l_y}\dagg a_{2,k,l_y-1}
  	  	+e^{ik/4}a_{2,k,l_y}\dagg a_{3,k,l_y+1} \right]
  	  +\text{H.c.}\right\}
  	\end{aligned}
  	\label{eq:hamiltonian0}
  \end{equation}
  Note that we take $\hbar=1$.
  Diagonalizing this matrix yields single-particle energy eigenstates with annihilation operator $b_{k,s}$, and a Hamiltonian $H_0=\sum_{k,s}\omega_{k,s}b_{k,s}\dagg b_{k,s}$.
  The resulting band structure is shown in \cref{fig:instabilities}.
  In our convention, the lowest bulk band has Chern number $\sgn D_z$, the middle bulk band $0$ and the top bulk band $-\sgn D_z$ (calculated, e.g., through the method described in Ref.~\onlinecite{Fukui2005}).
  Accordingly, there is one pair of edge modes in each of the bulk gaps, one right-moving localized at the lower edge and one left-moving at the upper.

  Including the anomalous terms obtained from a calculation based on the Fermi-Hubbard model at half filling yields the full Hamiltonian \cref{eq:full_hamiltonian}.
  By means of a Bogoliubov transformation we obtain the magnon band structure and the unstable states~\cite{BlaizotRipka1986}, which form the basis for \cref{fig:instabilities} and \cref{fig:edge_instability}. The inclusion of nonlinear damping yields \cref{eq:nonlinear_eom}, which has been used to calculate \cref{fig:current}.
  In the end, we calculate the current by evaluating the expectation value of the particle current or energy current operator, which can be obtained for a given bond from the continuity equation~\cite{BernevigBook2013}.
  The current across a certain cut of the system is obtained by summing the current operators for all the bonds that cross it. 
  As the system we study is inversion symmetric, the total current in the $x$ direction vanishes.
  In order to specifically find the edge current, we thus define a cut through half of the system, for example from the top edge to the middle.

\subsection*{Unstable modes in Bogoliubov-de Gennes equation}
We consider the full Hamiltonian
\begin{equation}
  H=H_0-H_{\text{rot}}+H_{\text{amp}}.
  \label{eq:Hfull}
\end{equation}
$H_0-H_{\text{rot}}$ gives rise to the band structure shown in \cref{fig:instabilities} above,
while $H_{\text{amp}}$ contains the anomalous terms.
The idea of this section is to calculate which modes in \cref{eq:Hfull} are unstable. Ideally, those should be the relevant edge modes, and only those. 
It turns out that this is possible in presence of linear damping.

Following Ref.~\onlinecite{Peano2016}, we define the vector $\ket{a_k}=(a_{k,1},a_{k,2},\dots,a_{k,N},a_{k_0-k,1}\dagg,\dots,a_{k_0-k,N}\dagg)^T$,
where the index combines the label $l_y$ and the site label in the unit cell and therefore runs from $1$ to $N=3W-2$.
The Hamiltonian can generically be written
\begin{equation}
  H=\sum_k\left[ a_{k,s}\dagg\mu_{k,ss'}a_{k,s'} +\frac{1}{2}\left(a\dagg_{k,s}\nu_{k,ss'}a\dagg_{-k,s'}+\text{H.c.}\right)\right]
  \label{eq:generic_ham}
\end{equation}
where $\mu_k$ originates from $H_0-H_{\text{rot}}$ and $\nu_k$ from $H_{\text{amp}}$.
This form makes it evident that $\mu_{k}=\mu_k\dagg$ and $\nu_k=\nu_k^T$.
The equation of motion for this vector can be found from the Hamiltonian above and is
\begin{equation}
  \frac{d}{dt}\ket{a_k}=-i\sigma_z h_k\ket{a_k},
  \label{eq:eom}
\end{equation}
with $\sigma_z=\diag(1,1,\dots,-1,-1,\dots)$ ($N$ ``$+1$''s and $N$ ``$-1$''s), with
\begin{equation}
  h_k=\mat{\mu_k & \nu_k \\ \nu_k\dagg & \mu_{-k}^T}.
\end{equation}
We can then solve the eigenvalue problem and find stable and unstable modes. 
Furthermore, we can find the time-evolution for operators in the Heisenberg picture from \cref{eq:eom}. It is simply
  $\ket{a_k(t)}=e^{-i\sigma_zh_kt}\ket{a_k(0)}$.

\subsection*{Steady state of nonlinear equations of motion}
We start from the equations of motion \eqref{eq:nonlinear_eom} given in the main text, repeated here for convenience
\begin{equation}
  i\mat{ \dot \alpha_k\\ \dot \alpha_{-k}^*}
  =\mat{\tilde\omega_k-i\frac{\gamma+\eta|\alpha_k|^2}{2}& \E \\ -\E & -\tilde\omega_{-k}-i\frac{\gamma+\eta|\alpha_k|^2}{2}}
  \mat{ \alpha_k\\ \alpha_{-k}^* }.
  \label{eq:two_mode_eom_Methods}
\end{equation}
In the steady state, $|\alpha_k|^2=\text{const.}$, so we use the ansatz $\alpha_k=\exp(i\Delta t)\bar\alpha$, and $\alpha_{-k}^*=\exp(i\Delta t)z\bar\alpha$ for some complex numbers $z$, $\bar\alpha$ and real frequency $\Delta$. 
As we are only interested in a narrow range of momenta, we expand the dispersion relation to second order, as in the main text.
The pump frequency is set to match the edge mode at $k=\pi$, i.e., $\Omega_0=2\omega_{\pi}$.
As a consequence, $\tilde\omega_{\pi+q}=q\omega_\pi'+q^2\omega_\pi''/2+\O(q^3)$.

If there is an instability, the solution $\bar\alpha=0$ is unstable. 
Assuming $\bar\alpha\neq0$ (thus $z\neq0$), and for $\Delta=q\omega_\pi'$, we find the set of equations
\begin{align}
	\frac{1}{2}q^2\omega_\pi''-\frac{i}{2}(\gamma+\eta|\alpha_k|^2)+z\E&=0,
  	\label{eq:nonlinear_EOM1}\\
	-\frac{1}{2}q^2\omega_\pi''-\frac{i}{2}(\gamma+\eta|z\alpha_k|^2)-\frac{\E}{z}&=0
	\label{eq:nonlinear_EOM2}.
\end{align}
Multiplying the second equation by $|z|^2$, and subtracting the complex conjugate of the resulting equation from the first equation, one can show that $|z|^2=1$. 
With this condition \cref{eq:nonlinear_EOM1,eq:nonlinear_EOM2} coincide,
such that we can solve them for the intensity
\begin{equation}
  |\alpha_k|^2=\frac{1}{\eta}\left( -2i\E z-iq^2\omega_\pi''-\gamma \right).
  \label{eq:alpha_SS}
\end{equation}
This equation has solutions if and only if $4\E^2\geq q^4(\omega_\pi'')^2+\gamma^2$, which coincides with the condition for the instability. 
If this condition is fulfilled, we have 
\begin{equation}
  |\alpha_{\pi+q}|^2=\frac{1}{\eta}\left( \sqrt{4\E^2-q^4(\omega_\pi'')^2}-\gamma \right).
  \label{eq:nonlinear_SS}
\end{equation}

The steady-state edge magnon current
\begin{equation}
  \begin{aligned}
  	J_{\mathrm{SS}}&=\int_{-\Lambda}^\Lambda\frac{dq}{2\pi}\,|\alpha_{\pi+q}|^2\nu_{\pi+q} \\ 
	=\frac{2\omega_\pi'\sqrt{2\E}}{3\pi\eta\sqrt{\omega_\pi''}}&\left\{ 2\E\, F\left[ \sin^{-1}\left( \frac{\Lambda}{\sqrt{2\E/\omega_\pi''}} \right),-1 \right]\right.\\&\left.-\gamma\sqrt[4]{1-\gamma^2/(4\E^2)}\right\}.
  \end{aligned}
\end{equation}
where $F(k,m)$ is the elliptic integral of the first kind.

\subsection*{Particle current operator}The particle current operator is obtained from the continuity equation for the number of magnons.
We have 
\begin{equation}
  \dot n_n-i[H_0,n_n]=\dot n_n-i\sum_m[h_m,n_n]=0,
  \label{eq:number_continuity_equation}
\end{equation}
where $h_n$ are local Hamiltonians defined through
\begin{equation}
  H_0
  =\sum_nh_n.
\end{equation}
The second term in \cref{eq:number_continuity_equation} can be interpreted as a sum of the particle currents from $n$ to the neighbouring sites $m$.

\bibliography{library}{}

\appendix
\clearpage
\onecolumngrid
  \begin{center}
	\textbf{\Large Supplementary Material: Topological Magnon Amplification}
  \end{center}
\section*{Supplementary Note 1: Effective spin Hamiltonian from Fermi-Hubbard model}
In order to support our qualitative analysis above, we derive the polarization tensor in a kagome TMI from a Fermi-Hubbard model at half filling, with an on-site Coulomb repulsion $U$ much larger than the hopping $t$, $t/U\ll1$.
As is well known, the low-energy physics can be described by perturbing around the Mott insulator state~\cite{MacDonald1988}.
A contribution to the polarization arises to third order in the hopping~\cite{Bulaevskii2008} (hopping around a triangle), which is derived in detail below.

  Following Zhu \emph{et al.}~\cite{Zhu2014}, we consider a one-band Hubbard model with SOC
  \begin{equation}
	H_{\mathrm{Hubbard}}=-\sum_{\langle ij\rangle }\left[ \vec c_i\dagg\left( \tau_{ij}+\vec d_{ij}\cdot\bm\sigma \right)\vec c_j+\text{H.c.} \right]+U\sum_{j}n_{j\up}n_{j\down},
	\label{eq:Hubbard Hamiltonian}
  \end{equation}
  with $\vec c_i=(c_{j\up},c_{j\down})^T$, and SOC vector $\vec d_{ij}=i\vec n_{ij}t_{ij}\sin(\theta_{ij})$.
  The parameterization of $\vec d$ in terms of a unit vector $\vec n_{ij}$ and an angle $\theta_{ij}$ will become useful later.

  Zhu \emph{et al.}\ \cite{Zhu2014} derive the effective low-energy spin Hamiltonian to second order in the hopping, which is
  \begin{equation}
	\tilde H_{\mathrm{eff}}=\sum_{\langle ij\rangle }\frac{4t_{ij}^2}{U}\vec S_i\dagg\J_{ij}\vec S_j,
	\label{eq:effective spin Hamiltonian}
  \end{equation}
  where $S_j^\nu=\tfrac12\vec c_j\dagg\cdot\sigma^\nu\cdot \vec c_j$, and $\J_{ij}$ is the exchange tensor pertaining to bond $\langle ij\rangle $ and can be written
  \begin{equation}
	\J_{ij}\vec S_j=\cos(2\theta_{ij})\vec S_j+\sin(2\theta_{ij})(\vec{S}_j\times\vec{n}_{ij})
	+2\sin^2(\theta_{ij})\vec n_{ij}(\vec{n}_{ij}\cdot\vec{S}_j).
	\label{eq:exchange_tensor}
  \end{equation}
  The three terms give rise to the isotropic Heisenberg interaction, to asymmetric, and to symmetric exchange anisotropy, respectively.

  \section*{Supplementary Note 2: Polarization tensor}
  The direction in which the polarization may point is constrained in the same way as the DM vector associated to each bond [\emph{cf.}~\cref{eq:bare_hamiltonian}].
The reflection symmetry around the plane orthogonal to each bond constrains the vector to lie in this symmetry plane.
In addition, the lattice is three-fold rotation symmetric, as well as inversion symmetric around lattice sites, such that the direction of one vector determines that of all others.
The precise direction and magnitude of the vector may be obtained, for example, from perturbation theory in the Fermi-Hubbard model at half filling, as we demonstrate below.
The results in the main text in principle only require that the anisotropic part of the polarization tensor is nonzero, which is allowed whenever bonds are not centres of inversion.

  Microscopically, the anisotropy is due to spin-orbit coupling (SOC).
  We follow Zhu \emph{et al.}~\cite{Zhu2014}, who derive the electric polarization as a third-order hopping process,
  which is the lowest-order relevant contribution.
  It is given through~\cite{Zhu2014}
  \begin{equation}
	\vec P_{ij}=\vec p_{0,ij}
	\left[ \vec S_i\cdot\J_{ij}\vec S_j\cos\theta_{ijk}+\vec n_{ijk}\cdot\vec S_i\times\J_{ij}\vec S_j\sin\theta_{ijk} \right]
	\label{eq:full_polarization_tensor}
  \end{equation}
  where $k$ is the third site in the loop and with
  \begin{equation}
	\vec p_{0,ij}\equiv8ea\frac{t_{ij}t_{jk}t_{ki}}{U^3}(\vec e_{jk}-\vec e_{ki})=8ea\frac{t_{ij}t_{jk}t_{ki}}{U^3}(2\bm\rho_k-\bm\rho_i-\bm\rho_j).
  \end{equation}
  The vector $\vec p_{0,ij}$ points into the triangle, orthogonal to the bond $\langle ij\rangle $ and in the plane of the triangle.
  Importantly, this means that, when following bonds along a straight line, their polarization changes sign from bond to bond.

  The angle $\theta$ parametrizes the relative strength of the SOC.
  To first order in $\theta$, the only scalar quantity one can construct with one vector are of the form $\vec n\cdot(\vec S\times\vec S)$,
  which does not have the form we are interested in.
  Hence we expand to second order
  \begin{equation}
	\vec P_{ij}=\vec P_{ij}^{(0)}+\vec P_{ij}^{(1)}+\vec P_{ij}^{(2)}+\O(\theta^3),
  \end{equation}
  with
  \begin{subequations}
	\begin{align}
	  \vec P_{ij}^{(0)}&=\vec p_{0,ij}\vec S_i \cdot\vec S_j,\\
	  \vec P_{ij}^{(1)}&=\vec p_{0,ij}\left( 2\theta_{ij}\vec n_{ij}+\theta_{ijk}\vec n_{ijk} \right)\cdot
	  (\vec S_i \times\vec S_j),\\
	  \vec P_{ij}^{(2)}&=\vec p_{0,ij}\left\{ -(\vec S_i \cdot\vec S_j)[2\theta_{ij}^2+2\theta_{ij}\theta_{ijk}(\vec n_{ijk} \cdot\vec n_{ij})+\tfrac12\theta_{ijk}^2]
	  +2\theta_{ij}^2(\vec S_i \cdot\vec n_{ij})(\vec S_j \cdot\vec n_{ij})
	  +2\theta_{ij}\theta_{ijk}(\vec S_i \cdot\vec n_{ij})(\vec S_j \cdot\vec n_{ijk})\right\}.
	\end{align}
	\label{eq:polarization_by_order}
  \end{subequations}
  Physically, in the original Hubbard Hamiltonian, only the SOC term can generate spin flips, which is why we need to go to second order in the SOC to obtain anomalous pairing terms that generate two magnons from one photon.
  In the spin wave picture, terms such as $(\vec S_i \cdot\vec n_{ij})(\vec S_j \cdot\vec n_{ij})$ 
  and $(\vec S_i \cdot\vec n_{ij})(\vec S_j \cdot\vec n_{ijk})$ can lead to anomalous terms, which can lead to instabilities and thus amplification.
  In order to make progress, we need to apply the general model \cref{eq:effective spin Hamiltonian} to our particular problem.
  Note that while \cref{eq:effective spin Hamiltonian} predicts a positive $J$, experiments show that $J$ is in fact negative.
  This is a result of other contributions, such as exchange. 
  Thus, the measured $J$ cannot be used to determine the angle $\theta_{ij}$ in \cref{eq:effective spin Hamiltonian}.
  Instead, the angle needs to be fitted independently, or determined from a measurement of the spin-orbit effect.
  Comparing \cref{eq:effective spin Hamiltonian} to \cref{eq:bare_hamiltonian}, we can identify
  \begin{equation}
	\frac{4t_{ij}^2}{U}\cos(2\theta_{ij})=J_{\mathrm{SE}},\qquad
	\frac{4t_{ij}^2}{U}\sin(2\theta_{ij})\vec n_{ij}=\vec D_{ij},
	\label{eq:model_map}
  \end{equation}
  where $\vec D_{ij}$ is the vector in \cref{eq:bare_hamiltonian},
  $\tan(2\theta_{ij})=-|\vec D_{ij}|/J_{\mathrm{SE},ij}$
  quantifies the strength of the DM interaction relative to the Heisenberg coupling from superexchange $J_{\mathrm{SE}}$,
  and we have used the subscript SE to denote the superexchange contribution.
  In principle, all these quantities can differ from site to site, but here we study a translation-invariant Hamiltonian, which simplifies the description considerably.

  By lattice symmetry, $\vec D$ has to lie in the plane orthogonal to the bonds (since that is a symmetry plane).
  In the pyrochlore lattice, each bond is part of two triangles.
  The net DM interaction is the sum of the contribution from each triangle. 
  If we consider the corner-sharing cube that surrounds the tetrahedron, the DM vector lies in the plane of the cube face that also encompasses the bond, as derived for instance in Ref.~\onlinecite{Elhajal2005}.
  If we choose the upright triangles in \cref{fig:instabilities} to be part of tetrahedra pointing into the plane (and thus the upside-down triangles are part of tetrahedra pointing out of the plane),
  and consider the bond lying along $x$ in an upright triangle, we have $\vec n_{12}=-(\sqrt{2/3})\vec{\hat z}-1/\sqrt{3}\vec{\hat y}$ ($\vec{\hat z}$ points out of the plane, i.e., our coordinate system is right handed with \cref{fig:instabilities} being the $xy$-plane, with $x$ being horizontal and $y$ vertical).
  The DM vectors for the other bonds in the triangle can be obtained through rotation by $2\pi/3$ around $z$.
  The DM vectors in upside-down triangle then follow from reversing the vectors in the upright triangle ($\vec v\to-\vec v$).
  This argument assumes ordered bonds (here counterclockwise in all triangles).

  This determines $\vec n_{ij}$ and $\theta_{ij}\equiv\theta=(1/2)\tan^{-1}(D/J_{\mathrm{SE}})$.
  The spin-orbit contribution is assumed to be weak, such that $\theta$ is small.
  In analogy to a charged particle picking up a U(1) phase when hopping in a loop penetrated by a magnetic field, 
  $\theta_{ijk}$ and $\vec n_{ijk}$ parameterize the SU(2)-phase that is picked up by the electron spin when hopping around the triangle~\cite{Zhu2014}.
  Writing the hopping part of the original Hubbard Hamiltonian
  \begin{equation}
	H_t=-\sum_{\langle ij\rangle }\vec c_i\dagg\A_{ij}\vec c_j,
  \end{equation}
  we can identify
	$\A_{ij}\equiv\exp(i\theta_{ij}\vec n_{ij}\cdot\bm\sigma).$
  The lowest order contribution to the polarization comes from a third order hopping process around a triangle, during which an electric spin picks up the total rotation
  \begin{equation}
	\A_{ij}\A_{jk}\A_{ki}\equiv\exp(-i\theta_{ijk}\vec n_{ijk} \cdot\bm\sigma).
  \end{equation}
  This defines $\theta_{ijk}$ and $\vec n_{ijk}$.
  Due to translation and rotation symmetries, $\theta_{ijk}$ is the same for all bonds, and given through
  \begin{equation}
  	\begin{aligned}
	  \theta_{l,ijk}&=\cos^{-1}\left[ \frac{1}{8}\left( 3\cos(\theta)+5\cos(3\theta)-4\sqrt{2}\sin^2(\theta) \right) \right]\\
	  &=\sqrt{6}\theta+\O(\theta^2).
  	\end{aligned}
	\label{eq:thetaijk}
  \end{equation}
  The sign is ambiguous, and we have chosen $\theta_{ijk}>0$ in the second equality.
  The vector $\vec n_{ijk}$ depends on the bond we consider. For the bond connecting site 1 and 2 in the same unit cell (i.e., the lower edge in an upright triangle), we have 
  \begin{equation}
	  \vec n_{l,123}\propto \left( \sin^2(\theta)(1-2\sqrt{2}\cot(\theta)),
	  \frac{(2\sqrt{2}\cot(\theta)-1)\sin^2(\theta)}{\sqrt{3}},
	  \frac{5+7\cos(2\theta)+\sin(2\theta)/\sqrt{2})}{\sqrt{6}}\right )
	  \sim \hat{\vec z}+\O(\theta).
	\label{eq:n123}
  \end{equation}
  The vectors $\vec n_{231}$, $\vec n_{312}$ can be obtained from $\vec n_{123}$ through rotation by $2\pi/3$ and $4\pi/3$ around $z$.

  Terms such as $\vec S_i \cdot\vec S_j$, $S_i^zS_j^z$, $\vec S_i \times\vec S_j$ cannot change the angular momentum along $z$ and thus do not lead to anomalous terms.
  In the second order (in $\theta$) contribution to the polarization [\emph{cf.}~\cref{eq:polarization_by_order}], we have two promising terms.
  The second, however, yields
  \begin{equation}
	2\theta_{ij}\theta_{ijk}(\vec S_i \cdot\vec n_{ij})(\vec S_j \cdot\vec n_{ijk})
	=2\theta_{ij}\theta_{ijk}(\vec S_i \cdot\vec n_{ij})S_j^z+\O(\theta^3),
  \end{equation}
  and thus does not contribute to second order. 
  The remaining term is
  \begin{equation}
	  \vec P_{ij}^{(2)}=
	  \vec p_{0,ij}2\theta^2\vec S_i\cdot\left( \vec n_{ij}-\hat{\vec z}n_{ij}^z\right)
	  \vec S_j\cdot\left( \vec n_{ij}-\hat{\vec z}n_{ij}^z \right)+\cdots,
  \end{equation}
  where we have subtracted the component of the vector $\vec n_{ij}$ along $z$, because it does not lead to anomalous terms. 

  \section*{Supplementary None 3: Amplification Hamiltonian}
  Independent of whether we justify the existence of an anomalous term via symmetry considerations (\cref{sec:model}) or the microscopic derivation (Supplementary Note 2), the amplification Hamiltonian takes same form, due to symmetry constraints.
  Note that due to symmetry, $\vec P_{ij}$ is constrained to lie in the plane perpendicular to the bond.
  As the $z$ component is irrelevant, the resulting amplification Hamiltonian depends only on the modulus of the in-plane component of the polarization $\vec P$.
  Taking only the relevant term, the amplification Hamiltonian is written
  as (note the slightly odd convention where by addition of the indices, such as $m+n$ or $1+2$, we mean that we take the vectors to those sites and add them)
  \begin{equation}
	  H_{\text{amp}}
	  =-2\theta^2\sum_{\langle mn\rangle }\vec E_{\frac{m+n}{2}}\cdot \vec p_{0,mn}(\vec S_m \cdot\vec Q_{mn})(\vec S_n \cdot\vec Q_{mn}),
	\label{eq:Hamp}
  \end{equation}
  where $\vec Q_{mn}=\vec n_{mn}-\hat{\vec z}n_{mn}$ is perpendicular to the bond and points outside for upright triangles and inside for upside-down triangles.
  In fact, it is irrelevant whether $\vec Q_{mn}$ points in or out, since $\vec Q_{mn}\to-\vec Q_{mn}$ leaves \cref{eq:Hamp} unchanged.
  In the spin-wave picture, we have 
  $(\vec S_m \cdot\vec Q_{mn})(\vec S_n \cdot\vec Q_{mn})=S^-_mS^-_n(Q_{mn}^+)^2/4+\text{H.c.}+\cdots$, where $Q^\pm\equiv Q^x\pm iQ^y$.
  Since $\vec Q_{\mathrm{mn}}$ points out of the upward facing triangles (they are parallel to $\vec n_{mn}$),
  we have $Q^+_{12}=e^{i\pi/6}/\sqrt{3}$, $Q^+_{23}=e^{5i\pi/6}/\sqrt{3}$, and $Q^+_{31}=-i/\sqrt{3}$.
  We end up with
  \begin{equation}
	\begin{aligned}
	  H_{\text{amp}}&=-\frac{1}{4}\sum_{\langle mn\rangle }\vec E_{(m+n)/2} \cdot\vec p_{0,mn}\left[ a_ma_n(Q^+_{mn})^2+\text{H.c.} \right]\\
	  &=-\frac{1}{4}\sum_l\left[
		\vec p_{0,12}a_{1,l}\left( \vec E_{\frac{1+2}{2},l} a_{2,l}-
		\vec E_{\frac{1+2}{2},l-\frac{3}{2}}a_{2,l-3}\right)(Q^+_{12})^2
		\vec p_{0,23}a_{2,l}\left( \vec E_{\tfrac{2+3}{2},l} a_{3,l}-
		\vec E_{\frac{2+3}{2},l+\frac{3}{2}}a_{3,l+2}\right)(Q^+_{23})^2\right.\\&\qquad+\left.
		\vec p_{0,31}a_{3,l}\left( \vec E_{\tfrac{3+1}{2},l} a_{1,l}-
		\vec E_{\frac{3+1}{2},l-\frac{1}{2}}a_{1,l-1}\right)(Q^+_{31})^2+\text{H.c.}
	  \right]\\
	  &=-\sum_{k,l_y}\frac{\vec E_0}{12}e^{i\Omega_0t}
	  \left[
		\vec p_{0,31}e^{i\pi/3}a_{3,k,l_y}a_{1,-k,l_y}2i\sin\left( -k\delta_1/2\right)
		+\vec p_{0,23}e^{-i\pi/3}a_{2,k,l_y}\left( a_{3,-k,l_y}e^{ik\delta_2/2}-a_{3,-k,l_y+1}e^{-ik\delta_2/2} \right)\right.\\ &\qquad- \left.
		\vec p_{0,12}a_{1,k,l_y}\left( a_{2,-k,l_y}e^{-ik\delta_3/2}-a_{2,-k,l_y-1}e^{ik\delta_3/2} \right)
	  \right]+\text{H.c.},
	\end{aligned}
  \end{equation}
  where we have used $(Q_{12}^+)^2=\tfrac13e^{i\pi/3}$, $(Q_{23}^+)^2=\tfrac13e^{-i\pi/3}$, $(Q_{31}^+)^2=-\tfrac13$.
  The minus sign between the two terms in the round and square brackets above stems from the fact that the induced polarization switches sign going from a bond to an adjacent one.
  $\Omega_0$ is the frequency of the incoming radiation, $\vec E_0=\hat{\vec e}E_0$ its polarization and amplitude.
  If the radiation is polarized along $z$, at least to this order in perturbation theory, it has no effect on the TMI, thus we choose it to lay in the plane.

Recall $\vec p_{0,ij}=8ea\frac{t_{ij}t_{jk}t_{ki}}{U^3}(2\bm\rho_k-\bm\rho_i-\bm\rho_j)$ (where $i,j,k\in\{1,2,3\}$ and are all distinct).
Then $2\bm\rho_1-\bm\rho_2-\bm\rho_3=(3,-\sqrt{3})/4$, 
$2\bm\rho_2-\bm\rho_1-\bm\rho_3=(0,\sqrt{3})/2$ and $2\bm\rho_3-\bm\rho_1-\bm\rho_2=-(3,\sqrt{3})/4$.
We further define
\begin{equation}
  \E\equiv\frac{\sqrt{3}eat^3}{4U^3}|E_0|,
\end{equation}
proportional to the strength of the electric field,
and go into the rotating frame with respect to the Hamiltonian 
  $H_{\text{rot}}=\frac{\Omega_0}{2}\sum_{\alpha,l_y,p_x}a_{\alpha,p_x,l_y}\dagg a_{\alpha,p_x,l_y}.$
We arrive at
\begin{equation}
  \begin{aligned}
	H_{\text{amp}}&=-\E \sum_{k,l_y}\hat{\vec e}\cdot
	\left\{
	  \hat{\vec y}e^{i\pi/3}a_{3,k,l_y}a_{1,-k,l_y}2i\sin\left( -\delta_1k/2\right)
	  +\frac{\sqrt{3}\hat{\vec x}-\hat{\vec y}}{2} e^{-i\pi/3}a_{2,k,l_y}\left( a_{3,-k,l_y}e^{ik\delta_2/2}-a_{3,-k,l_y+1}e^{-ik\delta_2/2} \right)\right.\\ &\qquad\left.
	  +\frac{\sqrt{3}\hat{\vec x}+\hat{\vec y}}{2}a_{1,k,l_y}\left( a_{2,-k,l_y}e^{-ik\delta_3/2}-a_{2,-k,l_y-1}e^{ik\delta_3/2} \right)
	\right\}+\text{H.c.},
  \end{aligned}
  \label{eq:Hamp_written_out}
\end{equation}
From this form it is clear that the terms at $\pm k$ couple, so that it is we should combine negative and positive momenta.
Finally, choosing $\vec{\hat e}=\vec{\hat y}$, this leads to
\begin{equation}
  \begin{aligned}
	&H_{\text{amp}}=-\E\sum_{k>0,l_y}
	\left\{\vec a^T_{k,l_y}\mat{0&\frac{1}{2}e^{-\frac{ik\delta_3}{2}}&e^{\frac{i\pi}{3}}2i\sin(\frac{k\delta_1}{2})\\
	\frac{1}{2}e^{\frac{ik\delta_3}{2}}&0&-\frac{1}{2}e^{-\frac{i\pi}{3}+\frac{ik\delta_2}{2}}\\
	e^{\frac{i\pi}{3}}2i\sin\left( -\frac{k\delta_1}{2} \right)&-\frac{1}{2}e^{-\frac{i\pi}{3}-\frac{ik\delta_2}{2}}&0}
	\vec a_{-k,l_y}\right.\\
	&+\left.\frac{1}{2}\left( e^{\frac{ik\delta_2}{2}-\frac{i\pi}{3}}a_{3,k,l_y+1}a_{2,-k,l_y}+e^{-\frac{ik\delta_2}{2}-\frac{i\pi}{3}}a_{2,k,l_y}a_{3,-k,l_y+1}
	-e^{-\frac{ik\delta_3}{2}}a_{2,k,l_y-1}a_{1,-k,l_y}-e^{\frac{ik\delta_3}{2}}a_{1,k,l_y}a_{2,-k,l_y-1} \right)\right\}+\text{H.c.}
  \end{aligned}
\end{equation}

\section*{Supplementary Note 4: Amplification matrix element}
\begin{figure*}[tb]
  \centering
  \includegraphics[width=\linewidth]{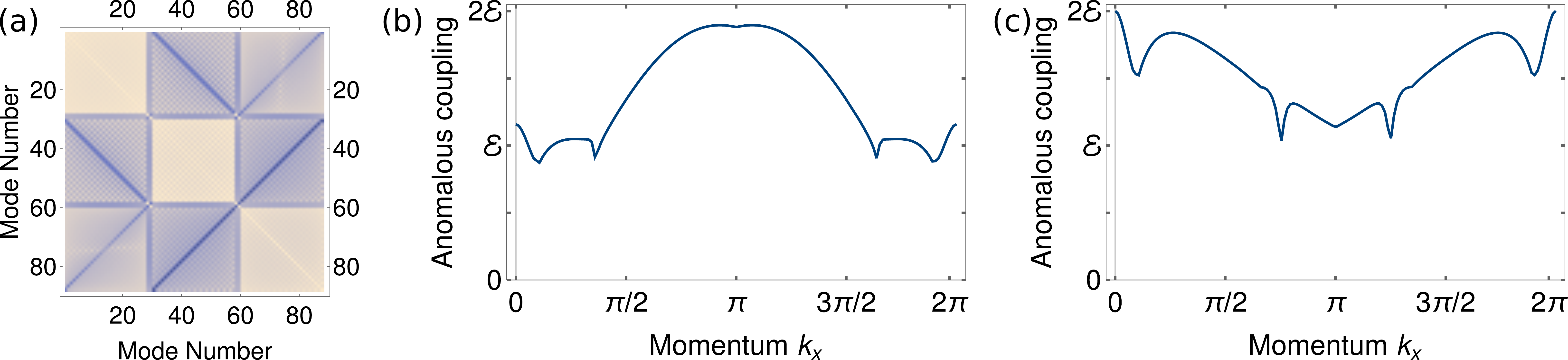}
  \caption{\textbf{Light-magnon matrix element.} (a) Modulus of the anomalous coupling between a given pair of modes at $k=\pi$. Dark blue corresponds to a maximum of $0.038J\approx2\E$, white to 0.
  (b) The maximum entry of the coupling matrix for wavevectors ranging from 0 to $2\pi$. A clear maximum arises around $k\approx\pi$.
  (c) The same plot, repeated for a polarization along $x$. In this case the anomalous coupling is suppressed for modes around $k=\pi$.
  Parameters are the same as \cref{fig:instabilities} (except for the polarization in the right plot).
  }
  \label{fig:anomalous_coupling}
\end{figure*}
In the section above we have derived the amplification Hamiltonian
\begin{equation}
  H_{\mathrm{amp}}=-\sum_{k,s,s'}\vec{E}_{0}\cdot\vec Q_{ss'}(k)a_{k,s}a_{-k,s'}+\text{H.c.}
  \label{eq:generic_amp}
\end{equation}
where here the generic indices $s,s'$ contain both the site label $\alpha$ and the unit cell label $l_y$. 

Diagonalizing the bilinear undriven Hamiltonian $H_0=\sum_{k}\vec a_k\dagg\matr\mu_k\vec a_k=\sum_k\vec b_k\dagg \matr\omega_{k}\vec b_k$,
where $\vec b_k=\matr U_k\dagg \vec a_k$ is a vector containing the annihilation operators of the energy eigenmodes and $\omega_{k}$ is a diagonal matrix.
Writing \cref{eq:generic_amp} in terms of energy eigenstates, we obtain
\begin{equation}
  H_{\mathrm{amp}}=-\sum_{k}\vec b_{k}\vec E_0\cdot(\matr U_k^*\vec Q(k)\matr U_k)\vec b_{k}.
\end{equation}
We can investigate the coupling strength between the various modes numerically, as is done in \cref{fig:anomalous_coupling}. 
The first conclusion, when considering the coupling matrix in the energy eigenbasis for wavevectors close to $\pi$ is that the anomalous coupling in between the edge modes is among the largest.
Comparable coupling strength is only achieved in between modes in differing bulk bands, as is seen from the diagonal lines in the off-diagonal blocks.
This can be appreciated by thinking about the form of the bulk wavefunctions along $y$, which are approximately standing waves with $0$ to $N_y-1$ nodes.
Since the matrix element between two bulk modes contains their product (with a constant applied field), summed over $y$, bulk modes with differing numbers of nodes approximately sum to zero. In between bands, the number of nodes within a unit cell changes, such that a full cancellation no longer occurs.

We next plot the maximum coupling strength between any of the modes as a function of wavevector.
From this plot we conclude that the anomalous coupling is most efficient around $k\approx\pi$.
This result can be understood to some degree by looking at the form of the amplification Hamiltonian \cref{eq:Hamp_written_out}.
Choosing the polarization of the applied field to lie along $y$, the first term coupling sites $1$ and $3$ is dominant.
In Fourier space this term has the functional form $\sin(k)$, such that it is largest around $\pi$, which roughly matches the shape in \cref{fig:anomalous_coupling}.
This conclusion is strongly dependent on the polarization we choose for the applied field.
We can plot the same quantities for a polarization along $x$, which turns off the coupling between sites 1 and 3.
In this case the maximum coupling strength no longer lies around $k=\pi$, which is plotted in \cref{fig:anomalous_coupling}.
Finally, this demonstrates one of the reasons why the agreement between the chiral waveguide model and the microscopic two-dimensional model is so good, namely that the matrix element is near unity (in units of $2\E$).

As we emphasize in the main text, the anomalous coupling strength is only one of the factors that influence whether a mode pair would become unstable under driving.
For example, all bulk mode pairs close to $k=\pi$ are far detuned in energy and thus cannot become unstable, regardless of the strength of their anomalous coupling. 

\end{document}